\documentstyle[aps,prl,twocolumn,graphicx]{revtex}
\begin{document}

\title{Full Current Statistics in Diffusive Normal-Superconductor Structures}

\author{W. Belzig\thanks{Present address: Department of Physics, University of
    Basel, Klingelbergstr. 82, 4056 Basel, Switzerland} and Yu. V. Nazarov }
\address{Department of Applied Physics and DIMES, TU Delft, Lorentzweg 1,
  2628 CJ Delft, The Netherlands\\{\rm Published: Phys. Rev. Lett. {\bf 87}, 067006 (2001)}}

\maketitle

\begin{abstract}
  We study the current statistics in normal diffusive conductors in
  contact with a superconductor. Using an extension of the Keldysh
  Green's function method we are able to find the full distribution of
  charge transfers for all temperatures and voltages. For the
  non-Gaussian regime, we show that the equilibrium current fluctuations
  are enhanced by the presence of the superconductor. We predict an
  enhancement of the nonequilibrium current noise for temperatures below
  and voltages of the order of the Thouless energy
  $E_{\text{Th}}=D/L^2$. Our calculation fully accounts for the
  proximity effect in the normal metal and agrees with experimental
  data. We demonstrate that the calculation of the full current
  statistics is in fact simpler than a concrete calculation of the
  noise. 
\end{abstract}

The electric current in conductors in general fluctuates. The full statistics
of these fluctuations can be accessed in some cases.\cite{levitov:96} However,
the full statistics is not easily accessible experimentally. The current
experiments mainly concentrate on noise power measurements. This has become an
important tool to extract information about the relevant charge transport
mechanism.\cite{blanter} One can, for example, extract the effective charge of
carriers responsible for the transport. Heterostructures with normal metals
(N) and superconductors (S) were intensively investigated. Generally, the
interest in these stems from the interplay between transport of doubly charged
Cooper pairs and singly charged normal quasiparticles.  The ratio between
noise power in the superconducting state and in the normal state can roughly
be interpreted as effective charge.  Experimental indications of a doubled
shot noise due to Andreev reflection in diffusive wires with one
superconducting lead have been reported in Refs.~\cite{sanquer,prober}.  Other
contributions report an enhancement of the current noise in SNS structures
\cite{strunk-klapwijk} with effective charges much larger that unity, possibly
originating from multiple Andreev reflections.

On the theoretical side first calculations of the noise \cite{khlus,dejong}
and the full statistics \cite{khmelnitzkii} for short contacts predicted an
enhanced shot noise with respect to the normal state value.  For certain
cases, such as tunnel junctions or disordered contacts, a doubled effective
charge was predicted. The drawback of these calculations is the limitation to
short contacts or small energies, since fully coherent propagation of
electrons and holes is assumed and the energy dependence of the scattering
amplitudes has been disregarded.  For the opposite regime, where coherence of
electrons and holes in the normal region plays no role, a modification of the
Boltzmann-Langevin approach has been put forward recently.\cite{nagaev}
Interestingly, it also gives a doubling of the shot noise in the incoherent
regime.  Similar limitations one finds in the available calculations of the
noise of SNS structures.\cite{sns-noise} It is tempting to ``interpolate''
between these two limits of coherent and incoherent propagation and to
conclude, that nothing interesting happens in the intermediate regime.  We
will show below, that this is not the case. As analogy let us note the
similarity to the linear conductance of a diffusive normal wire with one
superconducting lead.  Using the previously mentioned approaches one finds
that the conductance has exactly the same value as in the normal state
independent of temperature.  Only a full calculation using the Keldysh Green's
function technique revealed, that the conductance is significantly enhanced at
energies of the order of the Thouless energy $E_{\text{Th}}$\cite{yuli:96},
displaying a reentrant behaviour. This behaviour of the conductance has been
experimentally verified.  At present no calculation of current noise is
available, which fully accounts for the peculiarities of the proximity effect.

Another recent development is to study theoretically the full
statistical properties of current fluctuations. The method of choice to
do so is the so called {\em full counting statistics}, pioneered in
Ref.~\cite{levitov:96}. Introducing a counting field $\chi$ that couples
to the current operator, one can access the full current distribution.
Derivatives of the current with respect to $\chi$ immediately generate
all moments of the distribution. Thus, one circumvents the cumbersome
calculation of all the moments.

In this Letter we introduce a theoretical method that allows to obtain
the full current statistics for a wide class of SN-structures. We
further present concrete results for the equilibrium current
distribution and the nonequilibrium current noise with full account of
the proximity effect.  For simplicity, we restrict ourselves here to a
diffusive wire mounted between a normal reservoir and a
superconducting one.  First we determine the equilibrium current
distribution. It turns out that the large current fluctuations are more
probable in the superconducting case. Second we calculate the
temperature and voltage dependent current noise. We find that the noise
is enhanced (similar to the conductance) at voltages of the order of and
for temperatures below the Thouless energy. In the respective limits our
results agree with previous calculation.\cite{dejong,nagaev}

Let us first present the theoretical framework. We use a recently
developed extension \cite{yuli:99-2,belzig:00} of the Keldysh technique 
to compute the statistics of our proximity structure. There it was
shown that the current statistics can be obtained by imposing the
modified boundary condition on one reservoir's(L) Green's function
\begin{equation}
  \label{eq:bc1}
  \check G_{\text{L}}(\chi) = 
  e^{\frac{i}{2}\chi\check\tau_{\rm K}}
  \check G_{\text{L}}
  e^{-\frac{i}{2}\chi\check\tau_{\rm K}}\;.
\end{equation}
Here $\check G_{\text{L}}$ is the standard Green's function of a
reservoir and $\check\tau_{\rm K}=\hat\sigma_1\bar\tau_3$ is a matrix in
Keldysh($\hat{\ }$)-Nambu($\bar{\ }$) space.  Statistical properties are
encoded in the dependence on the counting field $\chi$. The current
probability distribution is found from
\begin{equation}
  \label{eq:pofn}
  P(I)=\int_{-\pi}^{\pi} d\chi e^{-S(\chi)-i\chi I t/e}\,.
\end{equation}
Here $t$ denotes the time of observation.  The action $S(\chi)$ can be
found from integration of the $\chi$-dependent current
\begin{equation}
  \label{eq:action}
  -i\frac{e}{t}\frac{\partial S(\chi)}{\partial\chi}=
  I(\chi)=\frac{1}{8e}\int dE \text{Tr}\left[ \check\tau_K \check
  I(\chi) \right]\,.
\end{equation}
The spectral matrix current in a diffusive wire is given by
\begin{equation}
  \label{eq:matrixcurrent}
  \check I(\chi) = -\sigma
  \check G(\chi)\frac{\partial}{\partial x}\check G(\chi)\,,
\end{equation}
where $\sigma$ is the conductivity. This matrix current has to be found by
defining an appropriate circuit.  The manual on how to do this can be found in
\cite{yuli:99-1}. Note, that the only change in comparison to the calculation
of the conductance is the modified boundary condition (\ref{eq:bc1}). All
other relations defining the actual circuit remain unchanged, as long as they
are respecting the full matrix structure.

We stress that our approach is not in contradiction with the general
scattering matrix approach of Ref.~\cite{khmelnitzkii}. If one would
know the electron-hole scattering amplitudes for the system under
consideration, and would not disregard their energy dependence, one
could obtain the same result. The characterization of these amplitudes
would have to be performed along the lines of Ref.~\cite{yuli:94}. We
also emphasize, that the calculation of the full current statistics can
be done in a finer, simpler and more compact way than a separate
calculation of it's second order perturbations series, i.~e. the noise.

\begin{figure}[htbp]
  \begin{center}
    \includegraphics[width=7cm,clip=true]{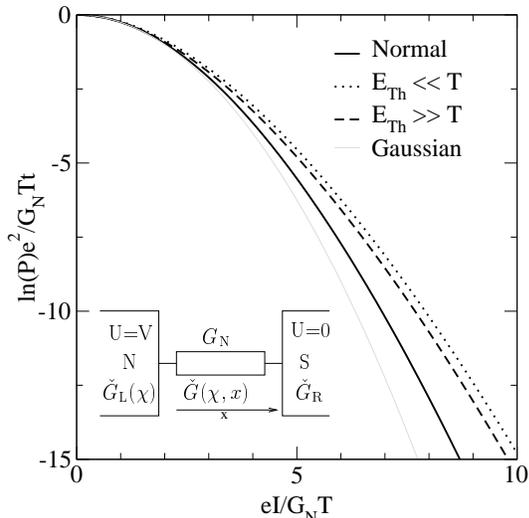}
    \caption[]{
      Equilibrium current distribution. The case, when the
      superconducting terminal is normal (solid line) does not depend on
      the Thouless energy.  In the superconducting state the
      distribution depends on $E_{\text{Th}}$. At low temperatures
      (dashed curve) large fluctuations are enhanced in comparison to
      the normal state.  This trend continues at temperatures above
      $E_{\text{Th}}$ (dotted curve). The deviations occur in the regime
      of non-Gaussian fluctuations. Note that the equilibrium noise is
      the same in all cases. The inset depicts our model system.}
    \label{fig:distr}
  \end{center}
\end{figure}

Let us now specify our system. A diffusive metal is connected to a
normal terminal on one end and to a superconducting terminal at the
other end.  Inside the mesoscopic wire the quasiclassical transport
equations are obeyed.\cite{eilenberger} In the normal metal
they read
\begin{equation}
  \label{eq:usadel}
  D \frac{\partial}{\partial x} 
  \left(\check G(x,\chi)\frac{\partial}{\partial x}\check G(x,\chi)\right) =
  \left[-iE\bar\tau_3\,,\,\check G(x,\chi)\right]\, .
\end{equation}
Here $D$ is the diffusion constant and $x$ the coordinate along the wire,
which has a length $L$.  It's conductance is $G_{\text{N}} = \sigma A/L$
(cross section $A$). At both ends boundary conditions to reservoirs have to be
supplied. At the normal end with ideal connection the Green's function is
continuous: $\check G(0,\chi)=\check G_{\text{L}}(\chi)$.  The other end is
connected to a superconducting reservoir by a contact of negligible
resistance, which leads to the boundary condition $\check G(L,\chi)=\check
G_{\text{R}}$.  A circuit representation of the system is depicted in the
inset of Fig.~\ref{fig:distr}.

In a normal reservoir (which we will consider in the rest of the paper)
$\check G_{\text{L}}$ is given by
\begin{equation}
  \label{eq:normalreservoir}
  \check G_{\text{L}}=\left(
    \begin{array}[c]{lr}
      \bar\tau_3 & \bar K\\
      0 & -\bar\tau_3 
    \end{array}
  \right)
  \,,\,
  \bar K=2\left(
    \begin{array}[c]{l}
      1-2f(E) \hfill 0 \\
      0 \quad 1-2f(-E)
    \end{array}
  \right)\,.
\end{equation}
The distribution function at voltage $V$ and temperature $T$ is given
by $f(E)=(\exp((E+eV)/T)+1)^{-1}$. A superconducting reservoir at zero
voltage is described by
\begin{equation}
  \label{eq:superreservoir}
  \check G_{\text{R}}=\left(
    \begin{array}[c]{cc}
      \bar R & 
      (\bar R-\bar A ) \tanh(E/2T)\\
      0 & \bar A 
    \end{array}
  \right)\,.
\end{equation}
Advanced and retarded Green's functions in (\ref{eq:superreservoir})
possess the structure $\bar R(\bar A) = g_{\text{R,A}}\bar\tau_3 +
f_{\text{R,A}}\bar\tau_1$, $f_{\text{R,A}} = i\Delta/((E\pm
i\delta)^2-\Delta^2)^{1/2}$ and $g_{\text{R,A}}$ following from the
normalization condition $f^2+g^2=1$ in a standard BCS-superconductor.

Let us briefly comment on the numerical procedure of the solution. It is
most convenient to solve the matrix equation (\ref{eq:usadel}) directly.
For this purpose the diffusive wire is represented by a discrete set of
$n$ nodes, each represented by a Green's function $\check G_k$ connected
in series by tunnel junctions of conductance
$g=(n+1)G_{\text{N}}$.\cite{yuli:99-1} The matrix current between two
neighboring nodes is then given by $ \check I_{k,k+1}=\frac{g}{2}
\left[\check G_k\,,\,\check G_{k+1} \right]$.  The right hand side of
(\ref{eq:usadel}) has a similar form and is represented as a
'decoherence' terminal (with a Green's function $\check
G_{\text{dec}}=-i 2(E/n E_{\text{Th}})\bar\tau_3$) connected to each
node.  The matrix current conservation for node $k$ follows from
discretizing Eq.~(\ref{eq:usadel}) and reads $[g(\check G_{k+1}+\check
G_{k-1})+\check G_{\text{dec}},\check G_k]=0$.  The resulting set of
equations for the nodes Green's functions is then solved by iteration.

\begin{figure}[htbp]
  \begin{center}
    \includegraphics[width=7cm,clip=true]{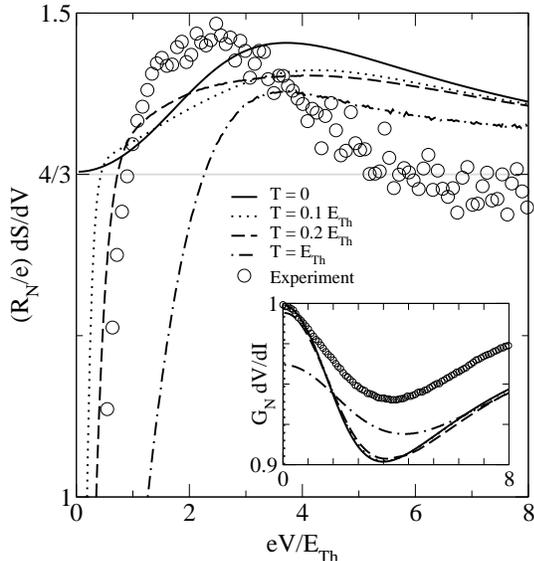}
    \caption[]{Differential noise and conductance. Temperature and
      voltages are in units of $E_{\text{Th}}$. The main plot shows our
      results for the differential noise together with experimental data
      from Kozhevnikov {\em et al.}\cite{prober} and the result of the
      Boltzmann-Langevin approach (grey lines). The inset shows the
      differential resistance.  Line styles and horiyontal axis
      correspond to the main plot, the small circles showing the
      corresponding experimental data of Kozhevnikov {\em et
        al.}\cite{prober} . At zero temperature, both differential noise
      and resistance show a reentrant behaviour, featuring a distinct
      bump at voltages of the order of $\approx 5E_{\text{Th}}$. A
      finite temperature tends to smear this signature of the proximity
      effect, but an enhancement above the Boltzmann-Langevin result is
      still clearly visible.  Comparison with the experimental data
      \cite{fit} show a good qualitative agreement for $T\approx 0.25
      E_{\text{Th}}$, which corresponds to the experimental value. Note
      that our results have no adjustable parameter.}
    \label{fig:noise}
  \end{center}
\end{figure}

Because of its importance in the following discussion, we have depicted the
differential resistance of the proximity wire in the inset of
Fig.\ref{fig:noise}. It displays the famous reentrance behaviour
\cite{yuli:96}, i.e. starting from the normal state resistance at zero voltage
a minimum at energies of the order of several $E_{\text{Th}}$ occurs. Above
the minimum it decays slowly to the normal state value as $\sim
(E_{\text{Th}}/E)^{1/2}$. Therefore, the Thouless energy is the central
quantity in the physics of the proximity effect. The disagreement of the
theory and the experimental data from Kozhevnikov {\em et al.}  \cite{prober}
may possibly result from heating effects, not accounted for in the theoretical
calculation.

As a first application, we study the distribution of current
fluctuations in equilibrium. We therefore put $V=0$ and find the
solution of the above equations for different values of $\chi$. We
evaluate the integral over $\chi$ in (\ref{eq:pofn}) in the saddle point
approximation, we take $\chi$ as complex and expand the exponent in
(\ref{eq:pofn}) around maximum. The integral yields then $ P(I) \approx
\exp(S(\chi)-\chi I t/e)$, which we plot implicitly as a function of
$I(\chi)$. To extract generic fluctuation properties of the proximity
effect we set here $\Delta\gg T,E_{\text{Th}}$.

Results of this calculation are displayed in Fig.~\ref{fig:distr}.  The
current $I$ is normalized by $G_{\text{N}}T/e$ and $\ln P(I)$ is plotted
in units of $G_{\text{N}}Tt/e^2$. The solid line shows the distribution
in the normal state, which does not depend on the Thouless energy. In
our units, this curve is consequently independent on temperature. In the
superconducting state the Thouless energy does matter, and the
distributions depend on the ratio $E_{\text{Th}}/T$.  We observe that
large fluctuations of the current in the superconducting case are
enhanced in comparison to the normal case, and in both cases are
enhanced in comparison to Gaussian noise.  For comparison we plotted the
Gaussian distribution $\sim \exp (- tI^2/4G_{\text{N}}T)$ of the current
measurements, owing to the fact that the conductance is the same in all
cases. The differences between the normal and the superconducting state
occur in the regime of non-Gaussian fluctuations.

Let us now turn to the nonequilibrium current noise as second application.
Previous results can be summarized in two statements. At temperatures larger
than an applied voltage the wire displays the usual thermal noise
$S_{\text{I}}(T\gg eV)=4 G_{\text{N}} T$ in accordance with the
fluctuation-dissipation theorem.  Note that in a calculation, which neglects
the proximity effect, the conductance of the normal wire is equal to the
normal state conductance $G_{\text{N}}$ independent of voltage or temperature.
For large voltages $eV\gg T$, on the other hand, the shot noise
$S_{\text{I}}(T\ll V)= (4/3) e G_{\text{N}} V$ is doubled in comparison to the
normal state value $(2/3) e G_{\text{N}} V$.  As discussed previously these
results do not depend on coherence between electrons and holes.  Consequently
the Thouless energy plays no role. However, this can not be true, since the
conductance is enhanced by approximately $10\%$ at energies of the order of
$E_{\text{Th}}$. In the following we will show, that the proximity effect
indeed changes the nonequilibrium noise of the system.

Using the method developed in \cite{yuli:99-2,belzig:00}, we are able to
take the coherence between electrons and holes fully into account. We
have to solve the Usadel equation (\ref{eq:usadel}) taking into account
the boundary condition (\ref{eq:bc1}). Then, the current noise power
is given by
\begin{equation}
  \label{eq:noise}
  S_{\text{I}}=2ei\left.\frac{\partial I(\chi)}{\partial \chi}\right|_{\chi=0}.
\end{equation}
There are two ways to attack this problem. One way to determine the noise
is to expand the Green's functions and Eq.~(\ref{eq:usadel}) to first order in
$\chi$ and thus to obtain an equation for the noise. The other way is to
solve the full matrix equations, and perform the differentiation in
(\ref{eq:noise}) numerically. Below we will use the second way to find
the nonequilibrium noise. Nevertheless, let us sketch the derivation of
an equation for the noise. We define $\check
G(x,\chi)=\check G_0(x)-i(\chi/2) \check G_1(x)$ and $\check
I(x,\chi)=\check I_0(x)-i(\chi/2) \check I_1(x)$. As result we find
\begin{eqnarray}
  \nonumber
  \check I_1(x) & = &
  - \sigma \left( \check G_0(x)\frac{\partial}{\partial x}\check G_1(x)
    + \check G_1(x)\frac{\partial}{\partial x}\check G_0(x)\right)\,, \\
  \label{eq:usadelnoise}
  \frac{D}{\sigma}\frac{\partial}{\partial x}\check I_1(x) & = &
  \left[-iE\bar\tau_3\,,\,\check G_1(x)\right]\,. 
\end{eqnarray}
From this equation the generalization of the Boltzmann-Langevin equation to
superconductors can in principle be derived. The boundary conditions at the
reservoirs read $\check G_1(0)=\left[\check\tau_{\text{K}},\check
  G_{\text{L}}\right]$ at the left end and $\check G_1(L)=0$ at the right end.
Finally the noise is $S_{\text{I}}=-e\int dE
\text{Tr}\check\tau_{\text{K}}\check I_1(x)$. By taking the trace of
Eq.~(\ref{eq:usadelnoise}) multiplied with $\check\tau_{\text{K}}$ it follows
that it does not matter, where the noise is evaluated, which is as it should be.

We turn now to concrete results for the noise power. The influence of the
proximity effect is most easily seen in the differential noise $dS/dV$.
Numerical results for different temperatures are displayed in
Fig.~\ref{fig:noise}. The inset shows the differential conductance for the
same parameters. The differential noise shows a remarkable enhancement at
energies of the order of the Thouless energy. Following a linear increase at
low voltage the differential noise overshoots the doubled normal differential
noise, which is recovered at large voltages. The maximal differential noise
occurs, if the voltage is of the order of $\approx 4E_{\text{Th}}$. The
largest enhancement is found for $T\ll E_{\text{Th}}$ and is roughly $10\%$.
At higher temperature the differential noise approaches the Boltzmann-Langevin
result \cite{nagaev}, shown as grey lines.  At zero temperature the reentrant
behaviour makes the connection to the result obtained within random matrix
theory.\cite{dejong} The nontrivial behaviour in the regime between these two
approaches shows the importance of phase coherence.

We can compare our results with experimental data on the noise power
obtained by Kozhevnikov {\em et al.}. \cite{prober,fit} The experimental
parameters correspond to $T/E_{\text{Th}}=0.2$. Given that our approach
contains no adjustable parameters, the agreement is very good. A
possible explanation for the difference between theory and experiment
is that heating effects in the experiments may be important. These have
been completely disregarded in the theoretical calculation.  This would
also explain the smaller reentrance of the differential resistance seen
in the experiment. Note, however, that the energy scale, at which the
influence of the proximity effect is seen, is unchanged.

In conclusion, we have developed a method to calculate statistical
properties (with emphasis on the current noise power) of
normal-metal--superconductor heterostructures. The method is embedded in
a matrix circuit theory \cite{yuli:99-1}, which allows to find the full
current statistics of a wide class of systems. We applied the method to
a normal diffusive wire with one normal and one superconducting
reservoir. In equilibrium it turns out that the current fluctuation
distribution differs from that of a purely normal system in the
non-Gaussian regime.  Large fluctuations of the current are enhanced by
the proximity effect.  For temperatures below $E_{\text{Th}}$ we found
that the nonequilibrium current noise shows a reentrant behaviour with a
maximum for voltages of the order of $E_{\text{Th}}$. This is in
qualitative agreement with experiment. \cite{prober}

We thank A.~A. Kozhevnikov and R.~J. Schoelkopf for discussions and for
providing us with the original data from Ref.~\cite{prober}.  W.B. was
financially supported by the ``Stichting voor Fundamenteel Onderzoek der
Materie'' (FOM) and the ``Alexander von Humboldt-Stiftung''.


\begin{thebibliography}{99}
\bibitem{fit} The difference to the plots presented in \cite{prober}
  stems from the fact, that we plot here the bare differential current
  noise $dS_{\text{I}}/dV$. That was obtained from the effective noise
  temperature (plotted in \cite{prober}) by differentiating
  $S_{\text{I}}(V)=T_{\text{N}}(V)/R(V)$.
\end{thebibliography}
\end{document}